\documentstyle[preprint,psfig,floats,pra,aps]{revtex}
\tightenlines
\begin{document} \draft

\title{\Large \bf Further Contents of Einstein's E = mc$^{2}$}

\author{Y. S. Kim\footnote{electronic mail: yskim@physics.umd.edu}}
\address{Department of Physics, University of Maryland, College Park,
Maryland 20742}

\maketitle

\begin{abstract}

The energy-mass content of Einstein's $E = mc^{2}$ is well known.
For a fixed value of mass, $E = mc^{2}$ is an energy-momentum relation
which takes the form $E = \sqrt{m^{2} + p^{2}}$.  This relation was
formulated in 1905 for point particles.  Since then, particles have
become more complicated.  They have internal space-time structures.
Massive particles carry the package of internal variables including
mass, spin and quarks, while massless particles have the package
containing helicity, gauge variables, and partons.  The question then
is whether these two different packages of variables can be unified
into one single covariant package as $E = mc^{2}$ does for the
energy-momentum relations for massive and massless particles.  The
answer to this question is YES.
\end{abstract}

\pacs{  }

\section{Introduction}\label{intro}

If the momentum of a particle is much smaller than its mass, the
energy-momentum relation is $E = p^{2}/2m + mc^{2}$.  If the momentum
is much larger than the mass, the relation is $E = cp$.  These two
different relations can be combined into one covariant formula
$E = \sqrt{m^{2} + p^{2}}$.  This aspect of Einstein's $E = mc^{2}$
is also well known.

In addition, particles have internal space-time variables.  Massive
particles have spins while massless particles have their helicities
and gauge variables.  Our first question is whether this aspect of
space-time variables can be unified into one covariant concept.
The answer to this question is Yes.  Wigner's little group does the
job.  In addition, particles can have space-time extensions.  For
instance, in the quark model, particles are bound states of quarks.
These bound states are called hadrons.  However, the hadrons appear
as a collection of partons when they move with speed close to the
velocity of light.  Quarks and partons seem to have quite distinct
properties.  In this report, we resolve this quark/parton puzzle.
We shall see that the quark model and parton model are two different
manifestations of the same covariant quantity.

By ``further contents'' of Einstein's $E = mc^{2}$, we mean that the
internal space-time structures of massive and massless particles can
be unified into one covariant package, as $E = \sqrt{m^{2} + p^{2}}$
does for the energy-momentum relation.  The mathematical framework of
this program was developed by Eugene Wigner in 1939~\cite{wig39}.
He constructed maximal subgroups of the Lorentz group whose
transformations will leave the four-momentum of a given particle
invariant.  These groups are known as Wigner's little groups.
Thus, the transformations of the little groups change the internal
space-time variables the particle.  The little group is a covariant
entity and takes different forms for the particles moving with
different speeds.

As for the relativistic extended particles, the most efficient
approach is to construct the representations of the little groups
using the wave functions which can be Lorentz-boosted.  This means
that we have to construct wave functions which are consistent with
all known rules of quantum mechanics.  It is possible to construct
harmonic oscillator wave functions which satisfy these conditions.
We can then take the low-speed and high-speed limits of the covariant
harmonic oscillator wave functions for the quark model and the
parton model  respectively.

The scope of this report is summarized in Table I.  We first use
the little groups to unify the spin variables for massive and massless
particles.  We then study the Lorentz-group contents of relativistic
extended hadrons to establish the quark-parton covariance.

%-------------------------------------------------------------------
\begin{table}%[h]

\caption{Further contents of Einstein's $E = mc^{2}$.  Massive and
massless particles have different energy-momentum relations.  Einstein's
special relativity gives one relation for both.  Wigner's little group
unifies the internal space-time symmetries for massive and massless
particles.  The quark model and the parton model can also be combined into
one covariant model.}

\vspace{3mm}

\begin{tabular}{rccc}
%\hline \\[-3.9mm]
%\hline
{}&{}&{}&{}\\
{} & Massive, Slow \hspace*{1mm} & COVARIANCE \hspace*{1mm}&
Massless, Fast \\[4mm]\hline
{}&{}&{}&{}\\
Energy- & {}  & Einstein's & {} \\
Momentum & $E = p^{2}/2m$ & $ E = [p^{2} + m^{2}]^{1/2}$ & $E = cp$
\\[4mm]\hline
{}&{}&{}&{}\\
Internal & $S_{3}$ & {}  &  $S_{3}$ \\[-1mm]
space-time &{} & Wigner's  & {} \\ [-1mm]
symmetry & $S_{1}, S_{2}$ & Little Group & Gauge
Transformations \\[4mm]\hline
{}&{}&{}&{}\\
Relativistic & {} & {} & {} \\[-1mm]
Extended & Quark Model & Covariant Model of Hadrons & Partons \\ [-1mm]
Particles & {} & {} & {} \\[2mm]
%\hline
%\hline
\end{tabular}
\end{table}
%----------------------------------------------------------------

In Sec. \ref{littleg}, we construct the little groups from their
definition that their transformations leave the four-momentum of a
given particle invariant.  In Sec. \ref{contrac}, we discuss in
detail how the little group for a massless particle can be obtained
as the zero-mass/infinite-momentum limit of the little group for the
massive particle.  The covariant oscillator formalism is spelled out
in detail in Sec. \ref{covham}.  In Sec. \ref{parton}, we use the
oscillator wave function to show that quarks and partons are
the same particles.

\section{Formulation of the Problem}\label{littleg}

The space-time symmetry of relativistic particles is dictated by
the Poincar\'e group~\cite{wig39}.  The Poincar\'e group is the group
of inhomogeneous Lorentz transformations, namely Lorentz transformations
preceded or followed by space-time translations.
Thus, the Poincar\'e group is a semi-direct product of
the Lorentz and translation groups.  The two Casimir operators of
this group correspond to the (mass)$^{2}$ and (spin)$^{2}$ of a given
particle.  Indeed, the particle mass and its spin magnitude are
Lorentz-invariant quantities.

The question then is how to construct the representations of the
Lorentz group which are relevant to physics.  For this purpose,
Wigner in 1939 studied the maximal subgroups of the Lorentz group
whose transformations leave the four-momentum of a given free
particle~\cite{wig39}.  These subgroups are called the little groups.
Since the little group leaves the four-momentum invariant, it governs
the internal space-time symmetries of relativistic particles.  Wigner
shows in his paper that the internal space-time symmetries of massive
and massless particles are dictated by the little groups which are
locally isomorphic to the three-dimensional rotation group and the
two-dimensional Euclidean groups respectively.

The group of Lorentz transformations consists of three boosts and
three rotations.  The rotations therefore constitute a subgroup of
the Lorentz group.  If a massive particle is at rest, its four-momentum
is invariant under rotations.  Thus the little group for a massive
particle at rest is the three-dimensional rotation group.  Then what is
affected by the rotation?  The answer to this question is very simple.
The particle in general has its spin.  The spin orientation is going
to be affected by the rotation!  If we use the four-vector coordinate
$(x, y, z, t)$, the four-momentum vector for the particle at rest is
$(0, 0, 0, m)$, and the three-dimensional rotation group leaves this
four-momentum invariant.  This little group is generated by
\begin{equation}\label{j3}
J_{1} = \pmatrix{0&0&0&0\cr0&0&-i&0\cr0&i&0&0\cr0&0&0&0} , \qquad
J_{2} = \pmatrix{0&0&i&0\cr0&0&0&0\cr-i&0&0&0\cr0&0&0&0} , \qquad
J_{3} = \pmatrix{0 & -i & 0 & 0 \cr i & 0 & 0 & 0
\cr 0 & 0 & 0 & 0 \cr 0 & 0 & 0 & 0} .
\end{equation}
These are essentially the generators of the three-dimensional rotation
group. They satisfy the commutation relations:
\begin{equation}\label{o3com}
[J_{i}, J_{j}] = i\epsilon_{ijk} J_{k} .
\end{equation}

If the rest-particle is boosted along the $z$ direction, it will pick
up a non-zero momentum component along the same direction.  The above
generators will also be boosted.  The boost will take the form of
conjugation by the boost matrix
\begin{equation}\label{boost}
B = \pmatrix{1&0&0&0\cr0&1&0&0\cr0&0 & \cosh\eta & \sinh\eta
\cr0 & 0 & \sinh\eta & \cosh\eta} .
\end{equation}
This boost will not change the commutation relations of Eq.(\ref{o3com})
for $O(3)$, and the boosted little group will still leave the
boosted four-momentum invariant.  Thus, the little group of a moving
massive particle is still  $O(3)$-like.

It is not possible to bring a massless particle to its rest frame.
In his 1939 paper~\cite{wig39}, Wigner observed that the little group
for a massless particle moving along the $z$ axis is generated by the
rotation generator around the $z$ axis, namely $J_{3}$ of Eq.(\ref{j3}),
and two other generators which take the form
\begin{equation}\label{n1n2}
N_{1} = \pmatrix{0 & 0 & -i & i \cr 0 & 0 & 0 & 0
\cr i & 0 & 0 & 0 \cr i & 0 & 0 & 0} ,  \quad
N_{2} = \pmatrix{0 & 0 & 0 & 0 \cr 0 & 0 & -i & i
\cr 0 & i & 0 & 0 \cr 0 & i & 0 & 0} .
\end{equation}
If we use $K_{i}$ for the boost generator along the i-th axis, these
matrices can be written as
\begin{equation}
N_{1} = K_{1} - J_{2} , \qquad N_{2} = K_{2} + J_{1} ,
\end{equation}
with
\begin{equation}
K_{1} = \pmatrix{0&0&0&i\cr0&0&0&0\cr0&0&0&0\cr i&0&0&0} , \qquad
K_{2} = \pmatrix{0&0&0&0\cr0&0&0&i\cr0&0&0&0\cr0&i&0&0} .
\end{equation}
The generators $J_{3}, N_{1}$ and $N_{2}$ satisfy the following set
of commutation relations.
\begin{equation}\label{e2lcom}
[N_{1}, N_{2}] = 0 , \qquad [J_{3}, N_{1}] = iN_{2} ,
\qquad [J_{3}, N_{2}] = -iN_{1} .
\end{equation}

In order to understand the mathematical basis of the above commutation
relations, let us consider transformations on a two-dimensional plane
with the $xy$ coordinate system.  We can then make rotations around
the origin and translations along the $x$ and $y$ directions.  If we
write these generators as $L, P_{x}$ and $P_{y}$ respectively, they
satisfy the commutation relations~\cite{knp86}
\begin{equation}\label{e2com}
[P_{x}, P_{y}] = 0 , \qquad [L, P_{x}] = iP_{y} ,
\qquad [L, P_{y}] = -iP_{x} .
\end{equation}
This is a closed set of commutation relations for the generators of the
$E(2)$ group.  If we replace $N_{1}$ and $N_{2}$ of Eq.(\ref{e2lcom})
by $P_{x}$ and $P_{y}$, and $J_{3}$ by $L$, the commutations relations
for the generators of the $E(2)$-like little group becomes those for
the $E(2)$-like little group.  This is precisely why we say that
the little group for massless particles are like $E(2)$.

It is not difficult to associate the rotation generator $J_{3}$ with
the helicity degree of freedom of the massless particle.   Then what
physical variable is associated with the $N_{1}$ and $N_{2}$
generators?  Indeed, Wigner was the one who discovered the existence
of these generators, but did not give any physical interpretation to
these translation-like generators.  For this reason, for many years,
only those representations with the zero-eigenvalues of the $N$
operators were thought to be physically meaningful
representations~\cite{wein64}.  It was not until 1971 when Janner
and Janssen reported that the transformations generated by these
operators are gauge transformations~\cite{janner71,kim97poz}.  The
role of this translation-like transformation has also been studied
for spin-1/2 particles, and it was concluded that the polarization
of neutrinos is due to gauge invariance~\cite{hks82,kim97min}.

\section{Contraction of O(3)-like to E(2)-like Little
Groups}\label{contrac}
The $O(3)$-like little group remains $O(3)$-like when the particle is
Lorentz-boosted.  Then, what happens when the particle speed becomes
the speed of light?  The energy-momentum relation
$E = \sqrt{m^{2} + p^{2}}$ become $E = p$.  Is there then a limiting
case of the $O(3)$-like little group?  Since those little groups are
like the three-dimensional rotation group and the two-dimensional
Euclidean group respectively, we are first interested in whether
$E(2)$ can be obtained from $O(3)$.  This will then give a clue to
obtaining the $E(2)$-like little group as a limiting case of
$O(3)$-like little group.  With this point in mind, let us look into
this geometrical problem.

In 1953, Inonu and Wigner formulated this problem as the contraction
of $O(3)$ to $E(2)$~\cite{inonu53}.  Let us see what they did.  We
always associate the three-dimensional rotation group with a spherical
surface.  Let us consider a circular area of radius 1 kilometer centered
on the north pole of the earth.  Since the radius of the earth is more
than 6,450 times longer, the circular region appears flat.  Thus, within
this region, we use the $E(2)$ symmetry group for this region.  The
validity of this approximation depends on the ratio of the two radii.

How about then the little groups which are isomorphic to $O(3)$ and
$E(2)$?  It is reasonable to expect that the $E(2)$-like little group
be obtained as a limiting case for of the $O(3)$-like little group
for massless particles.  In 1981, it was observed by Ferrara and Savoy
that this limiting process is the Lorentz boost~\cite{ferrara82}.
In 1983, using the same limiting process as that of Ferrara and Savoy,
Han {\it et al} showed that transverse rotation generators become the
generators of gauge transformations in the limit of infinite momentum
and/or zero mass~\cite{hks83pl}.

\begin{figure}[thb]
\centerline{\psfig{figure=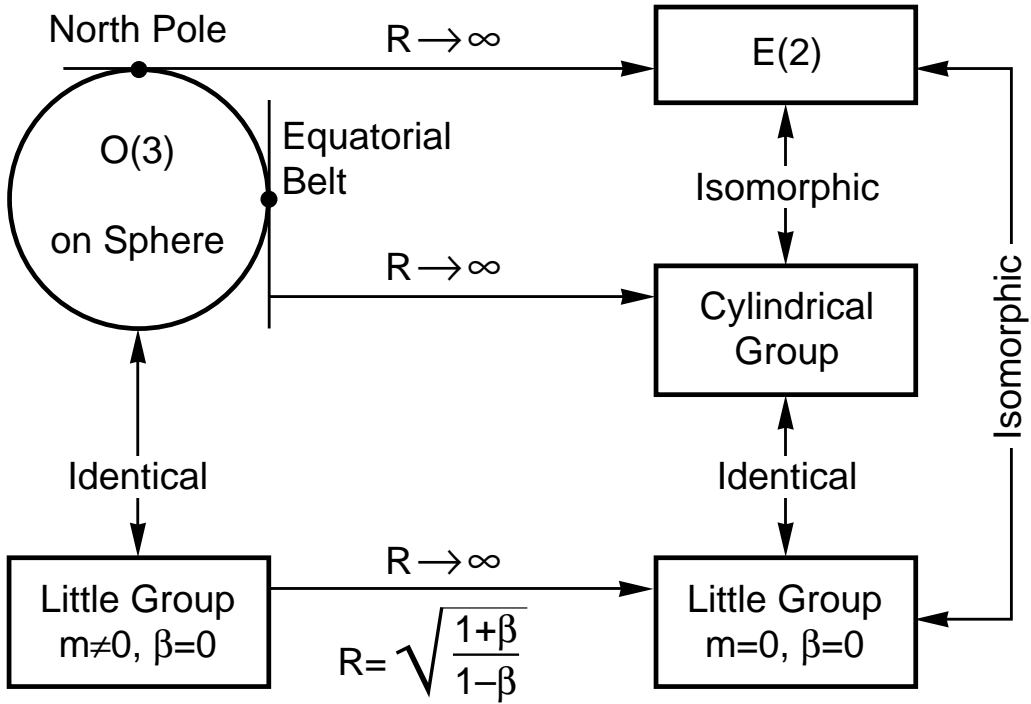,angle=0,height=100mm}}
\vspace{5mm}
\caption{Contraction of $O(3)$ to $E(2)$ and to the cylindrical group,
and contraction of the $O(3)$-like little group to the $E(2)$-like
little group.  The correspondence between $E(2)$ and the $E(2)$-like
little group is isomorphic but not identical.  The cylindrical group
is identical to the $E(2)$-like little group.  The Lorentz boost of
the $O(3)$-like little group for a massive particle is the same as
the contraction of $O(3)$ to the cylindrical group.}\label{f.isomor}
\end{figure}

Let us see how this happens. The $J_{3}$ operator of Eq.(\ref{j3}),
which generates rotations around the $z$ axis, is not affected by
the boost conjugation by the $B$ matrix of Eq.(\ref{boost}).  On
the other hand, the $J_{1}$ and $J_{2}$ matrices become
\begin{equation}
N_{1} = e^{-\eta} B^{-1} J_{2} B ,
\qquad N_{2} = -e^{-\eta} B^{-1} J_{1} B ,
\end{equation}
and they become $N_{1}$ and $N_{2}$ given in Eq.(\ref{n1n2}).  The
generators $N_{1}$ and $N_{2}$ are the contracted $J_{2}$ and $J_{1}$
respectively in the infinite-momentum/zero-mass limit.  In 1987, Kim
and Wigner studied this problem in more detail and showed that the
little group for massless particles is the cylindrical group which is
isomorphic to the $E(2)$ group~\cite{kiwi87jm}.  Their work is
summarized in Fig. \ref{f.isomor}.

This completes the second row in Table I, where Wigner's little group
unifies the internal space-time symmetries of massive and massless
particles.  The transverse components of the rotation generators become
generators of gauge transformations in the infinite-momentum/zero-mass
limit.

%---------------------------------------------------------------------------
%---------------------------------------------------------------------------

\section{Covariant Harmonic Oscillators}\label{covham}

We are now interested in constructing the third row in Table I.  As we
promised in Sec.~\ref{intro}, we will be dealing with hadrons which
are bound states of quarks with space-time extensions.  For this
purpose, we need a set of covariant wave functions consistent with the
existing laws of quantum mechanics, including of course the uncertainty
principle and probability interpretation.  The first wave function which
comes to our mind is the harmonic oscillator wave function.  If we are
interested in Lorentz-transforming them, the most straight-forward
method is to construct representations of the Poincar\'e group using
harmonic oscillators wave functions~\cite{dir45,yuka53,markov56,knp86}.

In this report, we start with the Lorentz-invariant differential
equation of Feynman, Kislinger, and Ravndal~\cite{fkr71}.  It is a
linear partial differential equation which has many different
solutions depending on boundary conditions.  Unlike in the case of
Feynman {\it et al}., we use normalizable wave functions which
constitute a representation of the $O(3)$-like little
group~\cite{knp86}.

Let us consider a bound state of two particles.  For convenience, we
shall call the bound state the hadron, and call its constituents quarks.
Then there is a Bohr-like radius measuring the space-like separation
between the quarks.  There is also a time-like separation between the
quarks, and this variable becomes mixed with the longitudinal spatial
separation as the hadron moves with a relativistic speed.  There are
no quantum excitations along the time-like direction.  On the other
hand, there is the time-energy uncertainty relation which allows
quantum transitions.  It is possible to accommodate these aspect within
the framework of the present form of quantum mechanics.  The uncertainty
relation between the time and energy variables is the c-number
relation~\cite{dir27}, which does not allow excitations along the
time-like coordinate.  We shall see that the covariant harmonic
oscillator formalism accommodates this narrow window in the present
form of quantum mechanics.

For a hadron consisting of two quarks, we can consider their space-time
positions $x_{a}$ and $x_{b}$, and use the variables
\begin{equation}
X = (x_{a} + x_{b})/2 , \qquad x = (x_{a} - x_{b})/2\sqrt{2} .
\end{equation}
The four-vector $X$ specifies where the hadron is located in space and
time, while the variable $x$ measures the space-time separation between
the quarks.  In the convention of Feynman {\it et al.}~\cite{fkr71},
the internal motion of the quarks bound by a harmonic oscillator
potential of unit strength can be described by the Lorentz-invariant
equation
\begin{equation}\label{osceq}
{1\over 2}\left\{x^{2}_{\mu} -
{\partial ^{2} \over \partial x_{\mu }^{2}}
\right\} \psi (x)= \lambda \psi (x) .
\end{equation}
It is now possible to construct a representation of the Poincar\'e group
from the solutions of the above differential equation~\cite{knp86}.

The coordinate $X$ is associated with the overall hadronic
four-momentum, and the space-time separation variable $x$ dictates
the internal space-time symmetry or the $O(3)$-like little group.  Thus,
we should construct the representation of the little group from the
solutions of the differential equation in Eq.(\ref{osceq}).  If the
hadron is at rest, we can separate the $t$ variable from the equation.
For this variable we can assign the ground-state wave function to
accommodate the c-number time-energy uncertainty relation~\cite{dir27}.
For the three space-like variables, we can solve the oscillator
equation in the spherical coordinate system with usual orbital and
radial excitations.  This will indeed constitute a representation of
the $O(3)$-like little group for each value of the mass.  The solution
should take the form
\begin{equation}
\psi (x,y,z,t) = \psi (x,y,z) \left({1\over \pi }\right)^{1/4}
\exp \left(-t^{2}/2 \right) ,
\end{equation}
where $\psi(x,y,z)$ is the wave function for the three-dimensional
oscillator with appropriate angular momentum quantum numbers.  Indeed,
the above wave function constitutes a representation of Wigner's
$O(3)$-like little group for a massive particle~\cite{knp86}.

Since the three-dimensional oscillator differential equation is
separable in both spherical and Cartesian coordinate systems,
$\psi(x,y,z)$ consists of Hermite polynomials of $x, y$, and $z$.
If the Lorentz boost is made along the $z$ direction, the $x$ and $y$
coordinates are not affected, and can be temporarily dropped from the wave
function.  The wave function of interest can be written as
\begin{equation}
\psi^{n}(z,t) = \left({1\over \pi }\right)^{1/4}\exp \pmatrix{-t^{2}/2}
\psi_{n}(z) ,
\end{equation}
with
\begin{equation}
\psi ^{n}(z) = \left({1 \over \pi n!2^{n}} \right)^{1/2} H_{n}(z)
\exp (-z^{2}/2) ,
\end{equation}
where $\psi ^{n}(z)$ is for the $n$-th excited oscillator state.
The full wave function $\psi ^{n}(z,t)$ is
\begin{equation}\label{2.6}
\psi ^{n}_{0}(z,t) = \left({1\over \pi n! 2^{n}}\right)^{1/2} H_{n}(z)
\exp \left\{-{1\over 2}\left(z^{2} + t^{2} \right) \right\} .
\end{equation}
The subscript $0$ means that the wave function is for the hadron at
rest.  The above expression is not Lorentz-invariant, and its
localization undergoes a Lorentz squeeze as the hadron moves along the
$z$ direction~\cite{knp86}.

It is convenient to use the light-cone variables to describe Lorentz
boosts.  The light-cone coordinate variables are
\begin{equation}
u = (z + t)/\sqrt{2} , \qquad v = (z - t)/\sqrt{2} .
\end{equation}
In terms of these variables, the Lorentz boost along the $z$
direction,
\begin{equation}
\pmatrix{z' \cr t'} = \pmatrix{\cosh \eta & \sinh \eta \cr
\sinh \eta & \cosh \eta } \pmatrix{z \cr t} ,
\end{equation}
takes the simple form
\begin{equation}\label{lorensq}
u' = e^{\eta } u , \qquad v' = e^{-\eta } v ,
\end{equation}
where $\eta $ is the boost parameter and is $\tanh ^{-1}(v/c)$.
Indeed, the $u$ variable becomes expanded while the $v$ variable becomes
contracted.  This is the squeeze mechanism illustrated discussed
extensively in the literature~\cite{kn73,knp91}.

The wave function of Eq.(\ref{2.6}) can be written as
\begin{equation}\label{10}
\psi ^{n}_{o}(z,t) = \psi ^{n}_{0}(z,t)
= \left({1 \over \pi n!2^{n}} \right)^{1/2} H_{n}\left((u + v)/\sqrt{2}
\right) \exp \left\{-{1\over 2} (u^{2} + v^{2}) \right\} .
\end{equation}
If the system is boosted, the wave function becomes
\begin{equation}\label{11}
\psi ^{n}_{\eta }(z,t) = \left({1 \over \pi n!2^{n}} \right)^{1/2}
H_{n} \left((e^{-\eta }u + e^{\eta }v)/\sqrt{2} \right)
\times \exp \left\{-{1\over 2}\left(e^{-2\eta }u^{2} +
e^{2\eta }v^{2}\right)\right\} .
\end{equation}

In both Eqs. (\ref{10}) and (\ref{11}), the localization property of
the wave function in the $u v$ plane is determined by the Gaussian factor,
and it is sufficient to study the ground state only for the essential
feature of the boundary condition.  The wave functions in Eq.(\ref{10})
and Eq.(\ref{11}) then respectively become
\begin{equation}\label{13}
\psi _{0}(z,t) = \left({1 \over \pi} \right)^{1/2}
\exp \left\{-{1\over 2} (u^{2} + v^{2}) \right\} .
\end{equation}
If the system is boosted, the wave function becomes
\begin{equation}\label{14}
\psi _{\eta }(z,t) = \left({1 \over \pi }\right)^{1/2}
\exp \left\{-{1\over 2}\left(e^{-2\eta }u^{2} +
e^{2\eta }v^{2}\right)\right\} .
\end{equation}
We note here that the transition from Eq.(\ref{13}) to Eq.(\ref{14})
is a squeeze transformation.  The wave function of Eq.(\ref{13}) is
distributed within a circular region in the $u v$ plane, and thus in
the $z t$ plane.  On the other hand, the wave function of Eq.(\ref{14})
is distributed in an elliptic region.  This is how the wave function is
Lorentz-boosted.

\section{Feynman's Parton Picture}\label{parton}

It is safe to believe that hadrons are quantum bound states of quarks having
localized probability distribution.  As in all bound-state cases, this
localization condition is responsible for the existence of discrete mass
spectra.  The most convincing evidence for this bound-state picture is the
hadronic mass spectra which are observed in high-energy
laboratories~\cite{knp86,fkr71}.  However, this picture of bound states
is applicable only to observers in the Lorentz frame in which the hadron
is at rest.  How would the hadrons appear
to observers in other Lorentz frames?

In 1969, Feynman observed that a fast-moving hadron can be regarded as a
collection of many ``partons'' whose properties do not appear to be
identical to those of quarks~\cite{fey69}.  For example, the number of
quarks inside a static proton is three, while the number of partons in a
rapidly moving proton appears to be infinite.  The question then is how
the proton looking like a bound state of quarks to one observer can appear
different to an observer in a different Lorentz frame?  Feynman made the
following systematic observations.

\begin{itemize}

\item[ a).] The picture is valid only for hadrons moving with velocity
     close to that of light.

\item[ b).] The interaction time between the quarks becomes dilated,
  and partons behave as free independent particles.

\item[ c).] The momentum distribution of partons becomes widespread as
  the hadron moves very fast.

\item[ d).] The number of partons seems to be infinite or much larger
  than that of quarks.

\end{itemize}

\noindent Because the hadron is believed to be a bound state of two or
three quarks, each of the above phenomena appears as a paradox,
particularly b) and c) together.  We would like to resolve this paradox
using the covariant harmonic oscillator formalism.

For this purpose, we need a momentum-energy wave function.  If the quarks
have the four-momenta $p_{a}$ and $p_{b}$, we can construct two independent
four-momentum variables~\cite{fkr71}
\begin{equation}
P = p_{a} + p_{b} , \qquad q = \sqrt{2}(p_{a} - p_{b}) .
\end{equation}
The four-momentum $P$ is the total four-momentum and is thus the hadronic
four-momentum.  $q$ measures the four-momentum separation between the quarks.

\begin{figure}[thb] %figure.4
\centerline{\psfig{figure=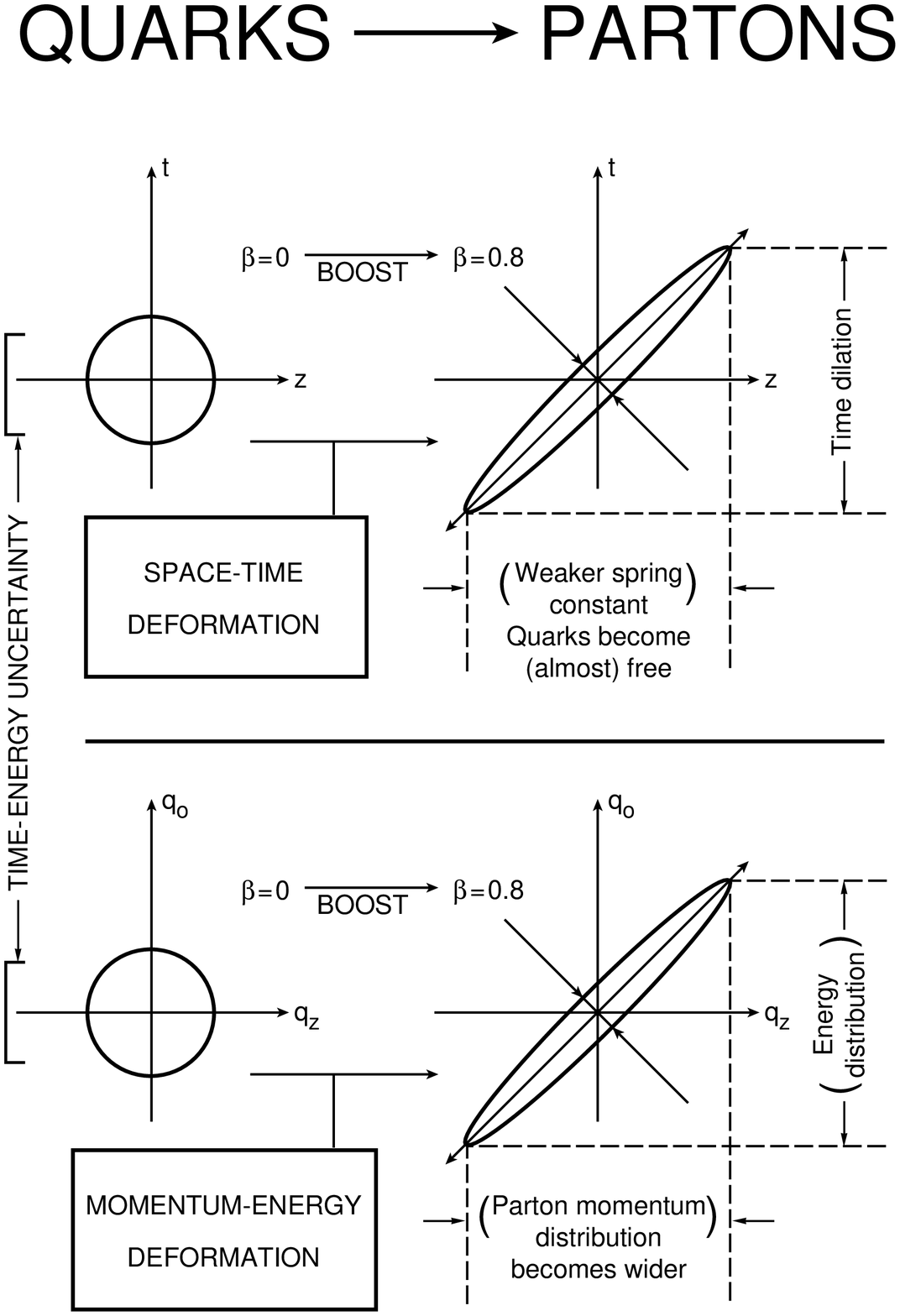,angle=0,height=120mm}}
\vspace{5mm}
\caption{Lorentz-squeezed space-time and momentum-energy wave functions.
As the hadron's speed approaches that of light, both wave functions
become concentrated along their respective positive light-cone axes.
These light-cone concentrations lead to Feynman's parton
picture.}\label{f.parton}
\end{figure}

We expect to get the momentum-energy wave function by taking the Fourier
transformation of Eq.(\ref{14}):
\begin{equation}\label{fourier}
\phi_{\eta }(q_{z},q_{0}) = \left({1 \over 2\pi }\right)
\int \psi_{\eta}(z, t) \exp{\left\{-i(q_{z}z - q_{0}t)\right\}} dx dt .
\end{equation}
Let us now define the momentum-energy variables in the light-cone coordinate
system as
\begin{equation}\label{conju}
q_{u} = (q_{0} - q_{z})/\sqrt{2} ,  \qquad
q_{v} = (q_{0} + q_{z})/\sqrt{2} .
\end{equation}
In terms of these variables, the Fourier transformation of
Eq.(\ref{fourier}) can be written as
\begin{equation}\label{fourier2}
\phi_{\eta }(q_{z},q_{0}) = \left({1 \over 2\pi }\right)
\int \psi_{\eta}(z, t) \exp{\left\{-i(q_{u} u + q_{v} v)\right\}} du dv .
\end{equation}
The resulting momentum-energy wave function is
\begin{equation}\label{phi}
\phi_{\eta }(q_{z},q_{0}) = \left({1 \over \pi }\right)^{1/2}
\exp\left\{-{1\over 2}\left(e^{-2\eta}q_{u}^{2} +
e^{2\eta}q_{v}^{2}\right)\right\} .
\end{equation}
Since we are using the harmonic oscillator, the mathematical form
of the above momentum-energy wave function is identical to that of the
space-time wave function.  The Lorentz squeeze properties of these wave
functions are also the same, as are indicated in Fig.~\ref{f.parton}.
These squeeze transformations perfectly consistent with the algorithms
of the Poincar\'e group~\cite{kim89}.

When the hadron is at rest with $\eta = 0$, both wave functions behave
like those for the static bound state of quarks.  As $\eta$ increases,
the wave functions become continuously squeezed until they become
concentrated along their respective positive light-cone axes.  Let us
look at the z-axis projection of the space-time wave function.  Indeed,
the width of the quark distribution increases as the hadronic speed
approaches that of the speed of light.  The position of each quark
appears widespread to the observer in the laboratory frame, and the
quarks appear like free particles.

%---------------------------------------------------------------------------
%---------------------------------------------------------------------------
Furthermore, interaction time of the quarks among themselves become
dilated.  Because the wave function becomes wide-spread, the distance
between one end of the harmonic oscillator well and the other end
increases as is indicated in Fig.~\ref{f.parton}.  This effect,
first noted by Feynman~\cite{fey69}, is universally observed in
high-energy hadronic experiments.  The period is oscillation is
increases like $e^{\eta}$.  On the other hand, the interaction time
with the external signal, since it is moving in the direction opposite
to the direction of the hadron, it travels along the negative
light-cone axis.  If the hadron contracts along the negative
light-cone axis, the interaction time decreases by $e^{-\eta}$.
The ratio of the interaction time to the oscillator period becomes
$e^{-2\eta}$.  The energy of each proton coming out of the Fermilab
accelerator is $900 GeV$.  This leads the ratio to $10^{-6}$.  This
is indeed a small number.  The external signal is not able to sense
the interaction of the quarks among themselves inside the hadron.

The momentum-energy wave function is just like the space-time wave
function.  The longitudinal momentum distribution becomes wide-spread
as the hadronic speed approaches the velocity of light.  This is in
contradiction with our expectation from nonrelativistic quantum
mechanics that the width of the momentum distribution is inversely
proportional to that of the position wave function.  Our expectation
is that if the quarks are free, they must have their sharply defined
momenta, not a wide-spread distribution.  This apparent
contradiction presents to us the following two fundamental questions:

\begin{itemize}

\item[a)].  If both the spatial and momentum distributions become
      widespread
      as the hadron moves, and if we insist on Heisenberg's uncertainty
      relation, is Planck's constant dependent on the hadronic velocity?

\item[b)].  Is this apparent contradiction related to another apparent
      contradiction that the number of partons is infinite while there
      are only two or three quarks inside the hadron?

\end{itemize}

The answer to the first question is ``No'', and that for the second
question is ``Yes''.  Let us answer the first question which is related
to the Lorentz invariance of Planck's constant.  If we take the product
of the width of the longitudinal momentum distribution and that of the
spatial distribution, we end up with the relation
\begin{equation}
<z^{2}><q_{z}^{2}> = (1/4)[\cosh(2\eta)]^{2}  .
\end{equation}
The right-hand side increases as the velocity parameter increases.
This could lead us to an erroneous conclusion that Planck's constant
becomes dependent on velocity.  This is not correct, because the
longitudinal momentum variable $q_{z}$ is no longer conjugate to the
longitudinal position variable when the hadron moves.

In order to maintain the Lorentz-invariance of the uncertainty product,
we have to work with a conjugate pair of variables whose product does
not depend on the velocity parameter.  Let us go back to Eq.(\ref{conju})
and Eq.(\ref{fourier2}).  It is quite clear that the light-cone variable
$u$ and $v$ are conjugate to $q_{u}$ and $q_{v}$ respectively.  It is
also clear that the distribution along the $q_{u}$ axis shrinks as the
$u$-axis distribution expands.  The exact calculation leads to
\begin{equation}
<u^{2}><q_{u}^{2}> = 1/4 , \qquad  <v^{2}><q_{v}^{2}> = 1/4  .
\end{equation}
Planck's constant is indeed Lorentz-invariant.

Let us next resolve the puzzle of why the number of partons appears to
be infinite while there are only a finite number of quarks inside the
hadron.  As the hadronic speed approaches the speed of light, both the
x and q distributions become concentrated along the positive light-cone
axis.  This means that the quarks also move with velocity very close
to that of light.  Quarks in this case behave like massless particles.

We then know from statistical mechanics that the number of massless
particles is not a conserved quantity.  For instance, in black-body
radiation, free light-like particles have a widespread momentum
distribution.  However, this does not contradict the known principles
of quantum mechanics, because the massless photons can be divided into
infinitely many massless particles with a continuous momentum
distribution.

Likewise, in the parton picture, massless free quarks have a wide-spread
momentum distribution.  They can appear as a distribution of an
infinite number of free particles.  These free massless particles are the
partons.  It is possible to measure this distribution in high-energy
laboratories, and it is also possible to calculate it using the covariant
harmonic oscillator formalism.  We are thus forced to compare these two
results.  Indeed, according to Hussar's calculation~\cite{hussar81},
the Lorentz-boosted oscillator wave function produces a reasonably
accurate parton distribution.

\section*{Concluding Remarks}
According to $E = mc^{2}$, the energy can be measured in kilograms.
For instance, Americans in the United States consume approximately
$300 Kg$ of electrical energy per year.  For a fixed value of mass,
the formula becomes $E = \sqrt{m^{2} + p^{2}}$, which unifies the
energy-momentum relation for massless particle and that for massive
particle with low speed.

We note that particles these days carry additional dynamical variables
and concept.  They carry internal space-time variables such as spin,
helicity, and gauge degree of freedom.  Wigner's little group unifies
all these variables into a single covariant regime.  In addition,
some particles, called hadrons, have their internal space-time
distributions.  These composite particles appear as two different
entities in quantum mechanics. We noted in this report that they
also can be unified.

\section*{Acknowledgments}
The author would like to thank Professor Victor Bashkov and the members
of the organizing committee for inviting him and for the hospitality
extended to him while in Kazan.  The citizens of Kazan were extremely
kind to him.

\end{document}